# Natural disasters and social entrepreneurship: An attention-based view

**Shihao Wei**
School of Management
Xi'an Jiaotong University
Xi'an, China
*shihaowei_xjtu@163.com*

**Christopher J. Boudreaux**
Department of Economics
Florida Atlantic University
777 Glades Road, OD 201B
Boca Raton, FL 33431 USA
*cboudreaux@fau.edu*

**Zhongfeng Su\***
School of Management
Xi'an Jiaotong University
Xi'an, China
*zhongfengsu@163.com*

**Zhan Wu**
Discipline of International Business
University of Sydney Business School
Sydney, Australia
*zhan.wu@sydney.edu.au*

forthcoming in *Small Business Economics*

October 2022 version

This is a pre-peer review version of the manuscript. The published version can be found at:
https://doi.org/10.1007/s11187-023-00822-x

Zhongfeng Su is the corresponding author.
Shihao and Zhongfeng appreciate the support for the National Natural Science Foundation of China (71832009).

**Abstract**

Drawing on the attention-based view, this study explores the joint effects of natural disaster intensity at the country-level with personal attributes in terms of gender, human capital, and fear of failure on the likelihood to enter social entrepreneurship. Using data on 107,386 observations across 30 countries, we find that natural disaster intensity has a positive effect on individuals' likelihood to engage in social entrepreneurship. In addition, the effect of natural disaster intensity is greater for males, individuals lacking human capital, and those who fear failure. Our study helps elaborate on the antecedents of social entrepreneurship and extends the consequences of natural disasters to entrepreneurship at the individual level.

**Plain English Summary**

This study reports that natural disaster intensity positively affects individuals' likelihood to engage in social entrepreneurship, and this effect is stronger for males, individuals lacking human capital, and those who fear failure. Therefore, instead of simply viewing natural disasters as a source of risk, individuals can leverage natural disasters to enter social entrepreneurship. In addition, the results also inform policymakers to encourage individuals to participate in social entrepreneurship to address and help alleviate the problems associated with natural disasters.

# 1 Introduction

Natural disasters result from dramatic climate change and extreme weather events like floods, earthquakes, volcanoes, slides, and windstorms (Huang et al., 2018). They severely affect people's lives and generate economic losses (Boudreaux et al., 2022b, 2022c; Vedula et al., 2022). In the last two decades, more than 11,000 natural disasters have caused about $2.56 trillion economic losses and took the lives of more than 450,000 people[1]. A substantial body of research demonstrates how natural disasters discourage economic growth (Cavallo et al., 2013; Klomp and Valckx, 2014; Loayza et al., 2012; Shabnam, 2014; Skidmore and Toya, 2002). In addition to the impact on the economic development, natural disasters also influence socioeconomic development through creating political instability and human conflict (Dell et al., 2014).

Yet, a growing body of literature suggests the relationship between natural disasters and entrepreneurship is more nuanced than prior studies acknowledge. On the one hand, natural disasters have devastating impacts on entrepreneurship activities. For instance, Boudreaux et al. (2019a) find that natural disaster intensity negatively affects new venture creation, measured by the number of newly registered per 1,000 people aged 15 to 64, primarily in the short term (1-2 years). This negative effect arises from increased uncertainty and risk (Monllor and Murphy, 2017). On the other hand, scholars note how natural disasters generate market inefficiencies and problematic social issues, creating opportunities for entrepreneurs as arbitrageurs and for new venture creation (Linnenluecke and Griffiths, 2010; Marino et al., 2008; Salvato et al., 2020). For example, Williams and Shepherd (2016a) report that natural disasters allow victims opportunities for new venture creation, which in turn, alleviates the negative shocks generated by natural disasters. Williams and Shepherd (2016b) also find that in the aftermath of the Haiti earthquake, entrepreneurial firms played key roles in reducing suffering. Muñoz et al. (2019) report that the continuous threats yielded by natural disasters facilitate entrepreneurs building the capability of preparedness, which aids them in rebuilding their new ventures or looking for new opportunities.

---
[1] www.germanwatch.org

Despite the importance of these studies, we still know little about how and when natural disasters encourage or discourage entrepreneurship. One explanation for the above ambiguity is that natural disasters exert heterogenous effects on alternative types of entrepreneurship. Specifically, as natural disasters cause destructive effects on economic activities and societal development (Boudreaux et al., 2022b, 2022c; Vedula et al., 2022), they are often perceived as severe threats and therefore can attract individuals' attention focus and affect their attention direction (Ghobadian et al., 2022; Pinkse and Gasbarro, 2019). For example, the devasting effects of natural disasters on infrastructure and the increased operation costs may direct persons' attention to uncertainty and risks associated with general entrepreneurial activities (i.e., corporate entrepreneurship, opportunity entrepreneurship, and necessity entrepreneurship), which inhibits their engagement in entrepreneurship. However, natural disasters also cause considerable and lasting human suffering and elicit severe threats to societal stability and development, thereby attracting individuals' attention to such social problems and providing opportunities for prosocial activities such as social entrepreneurship (SE). In support of this claim, we note the negative findings of Boudreaux et al. (2022b, 2022c, 2019a) are based on general entrepreneurship, but the positive findings of Williams and Shepherd (2016a, 2016b) are more similar to SE. Therefore, instead of taking entrepreneurship as a whole, it is imperative to investigate the effect of natural disasters on specific types of entrepreneurship to draw a more comprehensive and nuanced picture of the natural disasters and entrepreneurship linkage.

The purpose of our study is to examine the effects of natural disasters on SE. Drawing on the attention-based view (ABV), which claims that individuals' action is determined by their distribution of attention and "what decision-makers focus on, and what they do, depends on the particular context they are located in" (Ocasio, 1997, p. 189), we posit natural disasters are a key yet overlooked contextual factor that can attract individuals' attention to social problems and then affect their propensity to engage in SE (Dell et al., 2014; Doh et al., 2019). Moreover, as ABV highlights that external effects on individuals' attention allocation rely on personal attributes (Ocasio et al., 2018). Studies document individuals with diverse attributes vary in their concerns on social issues and respond differently (Cacciotti et al., 2016; Estrin et al., 2016; Jennings and Brush, 2013). As such, individuals with different attributes (e.g., gender, human capital, and fear of failure) may diversely allocate attention to social problems caused by natural disasters and then differ in their likelihoods to enter SE. Using data on 107,386 observations across 30 countries, we find that natural disaster intensity has a positive effect on an individual's likelihood to enter SE. Moreover, this positive effect is stronger for men, those with low human capital, and those who fear failure. Our findings are robust to instrumental variables, matching, alternative measures of our focal variable, and controls for country culture and resources.

Our findings make several contributions to the literature. First, our study contributes to theory on the natural disasters-entrepreneurship relationship. Drawing on ABV to examine how contextual and individual factors (e.g., gender, human capital, and fear of failure) jointly influence SE, our study introduces a cross-level framework providing a more integrated and nuanced picture on the antecedents of SE than prior studies (Saebi et al., 2019). Natural disasters serve as a key yet overlooked contextual factor attracting individuals' attention to social tragedies and their motivations to enter SE. In light of the damages caused by natural disasters, our study finds SE plays a pivotal role in addressing these social tragedies (Dell et al., 2014; Doh et al., 2019; Stephan et al., 2015). This finding extends earlier studies (Williams and Shepherd, 2016a, 2016b) by considering the role individual factors play in the natural disasters-SE relationship.

Second, we extend empirical findings in the literature on natural disasters and entrepreneurship. Although studies have advanced our knowledge of natural disasters and entrepreneurship, they were often conducted at the country or regional level (Boudreaux et al., 2022b, 2022c, 2019a). Few studies have examined the effects of natural disasters at the individual or firm-level (Huang et al., 2018). This is problematic because studies using data at the country or regional-levels of aggregation tend to mask entrepreneurs' responses at the individual and firm-level—a problem known as the *ecological fallacy,* i.e., drawing inferences about individual behavior from macro-level relationships (Freedman, 1999; Robinson, 1950; Seligson, 2002). To circumvent this issue, it is vital to examine the effects of natural disasters on individuals' behavior, such as the likelihood to enter SE.

Lastly, our study makes important contributions to entrepreneurship policy. Policymakers often focus on "picking winners" (Buffart et al., 2020; Lerner, 2010, 2009; Mazzucato, 2018; Shane, 2009) or identifying the institutional conditions with potential to increase employment and net job creation (Henrekson and Stenkula, 2010; Mason and Brown, 2013; Wennberg and Sandström, 2022). Yet, favoring foreign aid and multi-national organizations, policymakers have only paid limited attention to entrepreneurs' role in the aftermath of natural disasters. This is unfortunate because SE plays a vital role in disaster recovery. For example, in the aftermath of the Haiti earthquake, entrepreneurial firms played pivotal roles in reducing suffering (Williams and Shepherd, 2016a, 2016b). We extend these findings to inform policy by noting how individual-level attributes such as gender, education, and fear of failure influence individuals' attention toward SE in disasters' aftermath.

# 2 Literature review and hypotheses

*2.1 Attention Based View and SE*

The Attention Based View (ABV) highlights the role played by individuals' attention in decision making and behavior (Ocasio, 1997). On the one hand, ABV states individuals' actions depend on the particular context (Ocasio, 1997). The situation triggers the focus of individuals' attention, which makes contextual factors influence individual-level behavior (Dutt and Joseph, 2019; Kammerlander and Ganter, 2015; Nadkarni and Barr, 2008; Su et al., 2022; Sullivan, 2010). On the other hand, as individuals are "selective in the issues and answers they attend to" (Ocasio, 1997, p. 189), individuals with some attributes are more likely to allocate attention to certain issues than others (Bettinazzi and Zollo, 2022; Stevens et al., 2015; Tuggle et al., 2010). Thus, individuals' attributes affect one's attention, which in turn influence their behavior. ABV contends that contextual factors and personal attributes interact to affect the allocation of individuals' attention (Ocasio, 1997; Ocasio et al., 2018; Stevens et al., 2015). Consequently, in addition to examining the separate effects of contextual factors and personal attributes, ABV advocates combining them to draw an integrated picture of the antecedents of individuals' behavior (Stevens et al., 2015).

ABV has been widely used to explore the effects of several attention-related factors, especially those concerning crises/disasters and individual attributes on persons' attention distribution and subsequent actions. For instance, Ghobadian et al. (2022) and Pinkse and Gasbarro (2019) find that Covid-19 and climate change can influence firms' attentional processes to determine their focus of attention and the repertoire of adaptation measures they will adopt (Bettinazzi and Zollo, 2022; Stevens et al., 2015; Tuggle et al., 2010). Tuggle et al. (2010) find that CEO duality has a negative effect on board's allocation of attention to monitoring. Moreover, scholars also investigate the interactive effects of contextual and individual factors on attention distribution (Kammerlander and Ganter, 2015; Stevens et al., 2015). For example, Stevens et al (2015) show that firm prior performance and CEO's other-regarding values jointly affect their attention to social goals. Su et al. (2022) find that through broadening the scope of a TMT's attention, TMT heterogeneity has a positive impact on entrepreneurial bricolage, while competitive intensity weakens it. Hence, given the significance of ABV in illustrating how external contextual factors and individuals' attributes interactively affect attention allocation and behavior, we believe its insights are valuable and applicable to the cross-level phenomenon—SE (Kammerlander and Ganter, 2015; Saebi et al., 2019; Stevens et al., 2015).

Scholars have explored the drivers of SE from both the contextual- and individual-level. On the one hand, whether an individual will engage in SE depends on whether one directs attention to SE-related issues, which is influenced by the context (Ocasio, 1997). In other words, contextual factors influence individuals' likelihood to enter SE. Scholars have explored the effects of several contextual factors (Hoogendoorn, 2016; Hota, 2021). For instance, Pathak and Muralidharan (2018) find economic inequality has a positive effect on the likelihood to enter SE. Others have found government activism, postmaterialism, and socially supportive cultural norms positively affect SE (Stephan et al., 2015).

On the other hand, because individuals holding different attributes may direct their attention to different issues (Ocasio, 1997), personal attributes can influence the decision to become a social entrepreneur. Scholars have tested the effects of personal attributes and found several attributes matter. These attributes include human capital, social capital, and self-efficacy (Hota, 2021; Ruskin et al., 2016; Sahasranamam and Nandakumar, 2020). For instance, from a prosocial lens, Hockerts (2017) find that empathy, moral obligation, perceived social support, prior experience with social organizations, and social entrepreneurial self-efficacy all can increase an individual's social entrepreneurial intention.

Although these two branches of literature have been insightful, they have generally progressed in isolation (Saebi et al., 2019). This inhibits an integrated explanation for the antecedents of SE. Because ABV claims that individuals holding different attributes may direct their attention to different issues and respond to external stimuli differently (Ocasio, 1997, p. 189), personal attributes can influence the effects of external environments on the decision to become a social entrepreneur. Drawing on ABV, this study builds a cross-level model to investigate the effect of a key contextual factor—natural disasters on SE and how this relationship varies with individuals' attributes.

*2.2 Natural disaster intensity and SE*

As focusing events, natural disasters or crises often come to widespread public attention. For example, Haiti suffered a massive 7.0 magnitude earthquake in its largest metropolitan area in 2010. This disaster killed nearly 230 thousand people and left more than 1.5 million homeless. This attracted global attention, and the Haiti government received roughly $3.5 billion (U.S.) in donations by the end of 2010. Given the crucial role of natural disasters in attention distribution, scholars have adopted ABV to explore natural disasters' effects on firm behavior. For instance, because extreme weather events can damage production facilities or infrastructure, delay operations, disrupt supply chains, and change supply-demand conditions, Pinkse and Gasbarro (2019) suggest climate change can influence attentional processes (i.e., risk perception and perceived uncertainty of climate stimuli as well as the perceived impact of and past experience with climate stimuli) to affect firms' attention to routine or non-routine responses. Moreover, Ghobadian et al. (2022) find that while the Covid-19 pandemic-induced disruption can allocate

firms' attention to the Covid-19 pandemic, this attention distribution effect varies with the level of industry dynamism. Hence, natural disasters have a significant effect on individuals' attention allocation and subsequent actions.

We use natural disaster intensity, the number of people affected by natural disasters (the sum of those affected, injured, homeless, and killed), to explore the effect of natural disasters on the likelihood to join SE. Specifically, we posit natural disaster intensity positively affects individuals' likelihood to enter SE by directing individuals' attention to social problems. Natural disasters often create or exacerbate various social problems. For example, they not only damage public infrastructures such as roads, bridges, water and electricity supply, and communication services, but they also cause physiological and mental problems among affected inhabitants (Dutta, 2017). When natural disaster intensity is at a high level, a country suffers from many social problems and continuous threats caused by natural disasters (Muñoz et al., 2019; Williams and Shepherd, 2016a, 2016b). While governments, some international organizations, emergency relief agencies, charities, corporations, and other types of organizations often provide assistance for victims and affected areas, they cannot "meet all of the victims' critical needs and therefore suffering persists" (Shepherd and Williams, 2014, p. 953). In these instances, due to the problem-driven nature of attention (Dutt and Joseph, 2019; Kammerlander and Ganter, 2015; Nadkarni and Barr, 2008; Su et al., 2022; Sullivan, 2010), such unresolved social problems can generate external stimuli and attract individuals' attention. This, in turn, encourages them to direct attention to actions that can resolve the many social problems and achieve social goals—SE (Muñoz et al., 2019; Saebi et al., 2019). Therefore, individuals are more likely to engage in SE in a higher natural disaster intensity context. Conversely, when natural disaster intensity is at a low level, there are fewer social problems caused by natural disasters, and some can be addressed by governments or external institutions such as large charities, aid agencies, and multinational corporations (Dutta, 2017). That is, it becomes easier for governments and other organizations to solve social problems as natural disaster intensity decreases (Boudreaux et al., 2022b, 2022c, 2019a). Hence, few problems remain, which only generate trivial stimuli to individuals and play a less significant role in attracting their attention to social problems. Thus, individuals are less likely to engage in SE in a lower natural disaster intensity context.

*Hypothesis 1: Natural disaster intensity has a positive effect on individuals' likelihood to enter SE.*

*2.3 The joint effects of natural disaster intensity and personal attributes*

Although natural disaster intensity matters to individuals' likelihood to enter SE by attracting their attention to social problems, its effect varies among individuals with different attributes. This is because ABV states that individuals with varying attributes differ in their perceptions of external stimuli and their allocation of attention (Ocasio, 1997; Ocasio et al., 2018). To examine the effect of natural disaster intensity on individuals' likelihood to enter SE, it is vital to consider the role of personal attributes. Specifically, studies have identified individuals' gender, human capital, and fear of failure as key sources of variation in both individual attention allocation and the likelihood to pursue SE (Cacciotti et al., 2016; Cacciotti and Hayton, 2015; Estrin et al., 2016; Jennings and Brush, 2013; Sahasranamam and Nandakumar, 2020). Hence, we posit that gender, human capital, and fear of failure will matter for the way natural disaster intensity plays out for individuals' likelihood to enter SE. Figure 1 depicts this relationship.

**[Insert Figure 1 about here]**

More specifically, men and women tend to direct their attention to different issues that fit gender role stereotypes (Hechavarría and Brieger, 2022; Wood and Eagly, 2002). Women are often closely associated with prosocial-oriented attributes (i.e., other-regarding, empathy, care ethics, and compassion) and men tend to exhibit more achievement-orientated attributes (i.e., risk-taking, aggression, and confidence) (Hechavarría et al., 2017). Moreover, because women socialize more with friends, relatives, and colleagues, they have greater social embeddedness and stronger networks, which leads to more information (Lenz et al., 2021). Hence, the prosocial-oriented attributes and the social embeddedness lead women to pay more attention to social problems and prosocial activities than men (Dickel and Eckardt, 2021; Hechavarria et al., 2012; Hechavarría and Brieger, 2022). For example, Hechavarria et al. (2012) find the percentage of female founders in SE is larger than in commercial entrepreneurship, and Estrin et al. (2016) report that women give greater priority to SE over commercial entrepreneurship compared to men. However, scholars also find that males still account for a higher proportion of SE than females (Estrin et al., 2013; Rieger et al., 2021). There are two potential explanations. First, since women are involved with more household and family caretaking obligations (Lenz et al., 2021), these duties likely attract their attention and compete with their attention to social problems and prosocial activities, such as SE. Second, SE can sometimes be profit-motivated (Saebi et al., 2019). This is consistent with masculine achievement orientation that attracts their attention. In summary, men and women allocate different amounts of attention to social problems, which affects their entry into SE.

We posit the positive effect of natural disaster intensity on individuals' likelihood to enter SE is more noticeable for men. Natural disasters result in severe damage and cause considerable suffering to many (Dutta, 2017; Shepherd and Williams, 2014), thereby stimulating individuals to direct attention to these social problems (Dutta, 2017; Muñoz et al., 2019). Since men typically pay less attention to social problems (Dickel and Eckardt, 2021; Hechavarría and Brieger, 2022; Jennings and Brush, 2013), a natural disaster occurrence directs men's attention to social issues they might otherwise ignore. In other words, natural disasters play a greater role in complementing the lack of males' attention to social problems. Under such conditions, men can direct attention to not only social problems but also the financial rewards for engaging in SE. This will help increase their likelihood to enter SE. Conversely, as we mentioned above, while women typically pay more attention to social problems (Dickel and Eckardt, 2021; Hechavarría et al., 2017; Hechavarría and Brieger, 2022), their own personal and familial obligations can reduce their attention (Lenz et al., 2021). Although natural disasters generate additional social problems, they may only generate trivial stimuli and attract less attention. This is because social obligations may attract most of their attention. Hence, there is only limited space for the external stimuli generated by natural disasters to direct attention, meaning the effect of natural disaster intensity on individuals' likelihood to enter SE is more significant among men than women.

*Hypothesis 2: The positive effect of natural disaster intensity on individuals' likelihood to enter SE is stronger among men than women.*

Human capital is another important attribute affecting individuals' issue selections and subsequent attention distribution (Abramson and Inglehart, 1994; Estrin et al., 2016; Schofer and Fourcade-Gourinchas, 2001). Human capital is often acquired in higher education, but it can also be acquired through work experience (Becker, 2009; Kim and Li, 2014). When individuals are rich in human capital, they are more likely to realize the positive external effects of prosocial actions and develop prosocial attributes (Abramson and Inglehart, 1994; Estrin et al., 2016). This, in turn, directs their attention to social problems and encourages them to enter SE (Stephan et al., 2015). Studies have found a positive relationship between human capital and SE (Estrin et al., 2016; Sahasranamam and Nandakumar, 2020).

We posit the positive effect of natural disaster intensity on individuals' likelihood to enter

SE is more significant among individuals lacking human capital. Natural disasters create social problems that spur individuals to focus their attention and to take actions, such as SE, in response (Dutta, 2017; Muñoz et al., 2019; Shepherd and Williams, 2014). However, as attention is a valuable and scarce resource (Dutt and Joseph, 2019; Kammerlander and Ganter, 2015; Nadkarni and Barr, 2008; Ocasio, 1997, p. 189; Su et al., 2022; Sullivan, 2010), individuals rich in human capital can already use their knowledge base to distribute attention to some crucial yet "neglected problems in society involving positive externalities"(Santos, 2012, p. 342). Accordingly, most of their attention has been taken, which means that they are not necessarily distributing attention to the external stimuli generated by natural disasters. Hence, the additional positive role of natural disasters in attention allocation to social problems and SE may be weak. Conversely, individuals with less human capital often cannot realize the positive external effects of volunteering and prosocial actions (Abramson and Inglehart, 1994; Estrin et al., 2016). Therefore, these individuals are less likely to pay attention to social problems. In such instances, damages created by natural disasters to society and the suffering to victims can act as external stimulants to direct attention to social problems (Dutta, 2017; Muñoz et al., 2019; Shepherd and Williams, 2014). This is possible because the highly developed Internet and media provide more opportunities and channels for the government and social sectors to bring natural disasters to the top of the mind of individuals and make social problems caused by them more easily exposed to individuals who are lacking human capital. Hence, individuals lacking human capital are more stimulated by natural disasters to channel attention to social problems. This results in a stronger positive effect of natural disaster intensity on individuals' likelihood to engage in SE among them.

*Hypothesis 3: The positive effect of natural disaster intensity on individuals' likelihood to enter SE is stronger when they have below average human capital.*

Fear of failure represents the extent to which individuals are risk or loss averse (Cacciotti et al., 2016). When individuals have a high level of fear of failure, they often refuse to allocate attention to activities with a high failure rate, since they cannot endure the potential losses and negative emotions that come with failure (Morgan and Sisak, 2016; Wyrwich et al., 2016). For example, individuals who fear failures are less likely to distribute attention to social problems (Gonzales et al., 2017), since such social problems often exist in areas where governments and markets are ill-functioned and have high risks (Di Domenico et al., 2010). Studies report fear of failure lowers the likelihood of individuals becoming social entrepreneurs (Nicolás et al., 2018; Sahasranamam and Nandakumar, 2020).

We posit the positive effect of natural disaster intensity on individuals' likelihood to engage in SE is most salient for individuals with high fear of failure, suggesting the stimulation effect of natural disasters on individuals' attention to social problems is more significant for those who fear failure. When individuals fear failures, they are less prone to direct attention to social problems (Gonzales et al., 2017). However, the suffering and damages caused by natural disasters can incite prosocial traits such as empathy and compassion (Maki et al., 2019; Williams and Shepherd, 2016b). This natural disaster event and the traits brought out help assuage their fear of failure since they will be "punished less by stigma of failure", which encourage them to direct attention to social problems (Lee et al., 2022, p. 2016). Such effects can be stronger in disaster-affected areas, in that persons in the stricken areas have already lost so much and they have little to fail. That is, the stimuli generated by natural disasters help individuals overcome their fear of failure, which compensates for their lack of attention to social problems. This suggests the effect of natural disasters on individuals' attention to social problems is more significant when they fear failure. Conversely, because individuals who do not fear failure are less concerned with the high risks associated with social problems, they are already more likely to direct attention to social problems and then enter SE to address them. Thus, natural disasters play a smaller role in helping them overcome their fear of failure and directing their attention to social problems, weakening the effect of natural disaster intensity on their pursuit of SE.

*Hypothesis 4: The positive effect of natural disaster intensity on individuals' likelihood to enter SE is stronger when they fear failure.*

## 3 Methods

*3.1 Sample*

To test our hypotheses, we collected data from seven sources: the Global Entrepreneurship Monitor (GEM), the Center for Research on the Epidemiology of Disasters (CRED), the World Bank Development Index (WDI), the Human Development Report (HDR), the World Value Survey (WVS), the World Governance Index (WGI), and the Heritage Foundation's Economic Freedom Index (EFI). In particular, GEM surveys a portion of people across countries to capture their basic characteristics and entrepreneurship-related features annually, and the 2015 GEM wave took SE as a special topic. In light of "the broad and randomized nature of sampling in GEM . . . greatly improves the trustworthiness, generalizability, and repeatability" (Young et al., 2018, p. 417), GEM's 2015 wave data has been widely used in prior SE research (Estrin et al., 2016; Stephan et al., 2015). We gathered all individual-level data from GEM and matched them with country-level data, which we acquired from six other commonly used datasets, (Boudreaux et al., 2019b; Kim and Li, 2014; Stephan et al., 2015). Finally, we obtained 107,386 observations across 30 countries. Table 1 reports the sample's cross-country distribution.

**[Insert Table 1 about here]**

*3.2 Measures*

*3.2.1 Dependent variable*

*Social entrepreneurship (SE).* We gathered SE data from GEM's 2015 Social Entrepreneurship Special Topic. The questionnaire measures SE as, "Are you, alone or with others, currently trying to start or currently owning and managing any kind of activity, organization, or initiative that has a particularly social, environmental, or community objective?" We coded it "1" if the answer was a yes and "0" otherwise. The measure captures not only the nascent stage of SE but also new social enterprises. Hence, it provides a broader view of SE consistent with prior studies (Estrin et al., 2016; Hoogendoorn, 2016; Sahasranamam and Nandakumar, 2020; Stephan et al., 2015).

*3.2.2 Focal variable*

*Natural disaster intensity.* Following Boudreaux et al. (2022b), we measured natural disaster intensity using the number of people affected by natural disasters (the sum of those affected, injured, homeless, and killed), in various forms of natural disasters in one year. We obtained data from CRED. In addition, we used the number of affected and injured people as indicators of natural disaster intensity in robustness tests.

*3.2.3 Moderating variables*

*Gender.* We coded males as "1" and "0" for females. We gathered this variable from GEM.

*Human capital.* Since human capital is often acquired in higher education (Kim and Li, 2014), we coded human capital as "1" if the respondents have completed higher education and "0" otherwise. This measure is consistent with prior studies (Boudreaux and Nikolaev, 2019; De Clercq et al., 2013; Xavier-Oliveira et al., 2015). We gathered this variable from GEM.

*Fear of failure.* Following Estrin et al. (2016) and Pathak and Muralidharan (2018), we gathered data on fear of failure from GEM and coded it as "1" when the respondents indicate they are afraid of failures and "0" otherwise.

*3.2.4 Control variables*

We included control variables at both the country and individual levels. At the country level, we included seven variables. First, as economic development affects entrepreneurial activities, we controlled for GDP per capita (Audretsch et al., 2006; Boudreaux, 2019; Wennekers and Thurik, 1999). We gathered GDP per capita data from WDI. Second, the unemployment rate affects individuals' career choices (Estrin et al., 2016), such as unemployment can push individuals into entrepreneurship (Amit and Muller, 1995; Faria et al., 2010). We gathered data on the unemployment rate from WDI. Third, we controlled for human inequality, because it is a critical social issue that may trigger individuals' motivations to pursue SE (Boudreaux et al., 2022a; Bruton et al., 2021; Pathak and Muralidharan, 2018; Xavier-Oliveira et al., 2015). Human inequality is the mean values of inequality according to three areas—life expectancy, education, and income. We gathered human inequality data from the HDR[2]. Fourth, scholars have identified that postmaterialism—values like political freedom and participation, self-actualization, creativity, personal and social relationships, and self-actualization that extend beyond material needs—plays a significant role in SE (Stephan et al., 2015). We collected postmaterialism data from WVS. Fifth, given that the government can affect SE by enacting policies and determining resource allocation (Pathak and Muralidharan, 2018; Stephan et al., 2015), we controlled for three government-related factors in terms of governance quality, government spending, and government fiscal health. We acquired governance quality data from WGI and gathered data on government spending and government fiscal health from EFI. At the individual level, we include two control variables. The first one is age. Studies have found that age and its quadratic are closely associated with entrepreneurship (Kautonen et al., 2017; Lévesque and Minniti, 2011, 2006). That is, entrepreneurial entry increases until a certain age, after which it declines (Parker, 2018; Stephan et al., 2015). Second, we used individuals' entrepreneurship experience as a control variable, as it affects an individual willingness to engage in entrepreneurship (Estrin et al., 2016). Table 2 reports our variables' sources and measures.

---

[2] https://report.hdr.undp.org/.

**[Insert Table 2 about here]**

*3.3 Estimation methods*

We chose our estimation methods following several procedures. First, we combined the country- and individual-level data. Therefore, we use multilevel models to test our hypotheses, since they address the nested structure of our data and contribute to solving the issue of unobserved heterogeneity across countries (Autio et al., 2013; Boudreaux et al., 2019b; Hofmann et al., 2000; Pathak and Muralidharan, 2018). Second, given that our dependent variable is binary, we conducted multilevel mixed logistic analyses. Third, following existing studies (Boudreaux et al., 2022b; Estrin et al., 2016; Pathak and Muralidharan, 2018), we lagged country-level predictors, including the focal variable—natural disaster density, by one year to examine their effects on SE in the following year, which may help mitigate the issue of reverse causation. We calculated the variance inflation factor (VIF) to address potential multicollinearity issues. Our mean VIF is 2.47 and with a high of 6.77, suggesting no significant multicollinearity problem.

We conducted our analyses in four steps. First, we only included the control variables in the model. This serves as the baseline. Second, we added natural disaster intensity and all three individual attributes into the model. Third, we included the interactions of natural disaster intensity with gender, human capital, and fear of failure to examine their moderating effect, separately. Finally, we included all of the interaction terms in the full model.

## 4 Findings

*4.1 Results*

Table 3 presents the correlation matrix and descriptive statistics.

**[Insert Tables 3 and 4 about here]**

Table 4 displays the results of the multilevel mixed effects logistic analysis. Model 1.1 finds a positive effect of age ($\beta = 0.002$; $p = 0.881$) and a negative effect of age square ($\beta = -0.034$, $p = 0.005$) on SE, suggesting an inverted-U shaped linkage between age and the likelihood to enter SE. Both entrepreneurship experience ($\beta = 0.145$, $p = 0.000$) and human capital ($\beta = 0.278$, $p = 0.000$) have positive effects on SE but fear of failure has a negative effect ($\beta = -0.072$, $p = 0.000$). In addition, consistent with the findings of Estrin et al. (2013) and Rieger et al. (2021), Model 1.1 reports that males ($\beta = 0.115$, $p = 0.000$) are more likely to join SE. However, this finding is counterintuitive since most scholars contend women have a higher level of prosocial motivation than men (Dickel and Eckardt, 2021; Estrin et al., 2016).

In Model 1.2, we augment the baseline model to include our focal variable natural disaster intensity. We observe a positive effect of natural disaster intensity on the likelihood to engage in SE ($\beta = 0.477$, $p = 0.003$), providing support for Hypothesis 1. Model 1.3 includes the interaction of natural disasters intensity and gender. We observe a positive coefficient of the interaction term ($\beta = 0.034$, $p = 0.032$), suggesting the positive effect of natural disaster intensity on SE ($\beta = 0.474$; $p = 0.004$) becomes larger for men relative to women. Thus, we find evidence to support Hypothesis 2. Model 1.4 augments Model 1.2 to include the interaction of natural disaster intensity and human capital. We observe a negative coefficient of the interaction term ($\beta = -0.039$, $p = 0.006$), suggesting the positive effect of natural disaster intensity on SE ($\beta = 0.492$; $p = 0.003$) is weaker for individuals with a college education or higher (i.e., high human capital). This finding supports Hypothesis 3. Model 1.5 includes the interaction of natural disaster intensity and fear of failure. We observe a positive coefficient on the interaction term ($β = 0.051$, $p = 0.002$), suggesting the positive effect of natural disaster intensity on SE ($β = 0.480$; $p = 0.003$) is more significant for those who fear failure. This finding supports Hypothesis 4. Model 1.6, the full model, includes all of three interaction terms, and exhibits the same result.

To gain a better understanding of effect sizes, we plot the moderating effects in Figures 2, 3, and 4, respectively. Figure 2 reports the interaction between natural disaster intensity and gender. The results reveal a positive effect of natural disaster intensity on the likelihood to enter SE. This effect is stronger for men relative to women. Figure 3 reports the interaction between natural disaster intensity and human capital. The results reveal a positive effect of natural disaster intensity on the likelihood to enter SE, and this effect is stronger for individuals with less than a college education (i.e., low human capital). Figure 4 exhibits the interaction between natural disaster intensity and fear of failure. The results reveal a positive effect of natural disaster intensity on the likelihood to enter SE, and the effect is stronger for individuals who fear failure.

**[Insert Figure 2-4 about here]**

*4.2 Robustness tests*

*4.2.1 Instrumental variable analysis*

To examine the robustness of our findings, we conducted several additional tests. First, to resolve the potential issues of omitted variables and reverse causation, we adopted the instrumental variable method to re-analyze our models. Following Huang et al. (2018), we chose population density as the instrument, because it is closely related to natural disaster intensity yet uncorrelated with SE. We collected data on a country's population density from WDI, measured as the number of people per square kilometer. We conducted the instrumental variable analysis in two steps. In the first stage, we regressed natural disaster intensity on population density and included all control variables in the model. Table 5 reports the results from this first stage regression in column 1. We observe a negative and statistically significant coefficient of population density ($β = -0.112$, $p = 0.000$), which provides support for the instrumental relevance condition. In the second stage, we included the fitted values of natural disaster intensity acquired from the first-stage regression. The results are qualitatively similar to our main results. In Model 2.5, we observe disaster intensity has a positive effect on SE ($β = 0.712$, $p = 0.000$), and this effect is larger for men relative to women ($β = 0.216$, $p = 0.028$), for those with low human capital ($β = -0.480$, $p = 0.000$), and for those who are afraid of failure ($β = 0.273$, $p = 0.004$). These findings provide additional support for our hypotheses.

**[Insert Table 5 about here]**

*4.2.2 Additional controls for culture*

Another concern is that national culture plays an important role in influencing individuals' pursuit of SE (Hechavarría and Brieger, 2022; Hota, 2021). As such, omitting this variable could potentially bias our parameter estimates. Therefore, we included Hofstede's culture index in our models to control for its potential effect (Hofstede, 2001). We report these findings in the supplemental appendix Table A1. According to Model 3.6, we observe negative coefficients for power distance ($β = -0.019$, $p = 0.031$), long-term orientation ($β = -0.020$, $p = 0.006$), and indulgence versus restraint ($β = -0.025$, $p = 0.002$). In contrast, we observe positive coefficients for masculinity ($β = 0.041$, $p = 0.000$) and uncertainty avoidance ($β = 0.013$, $p = 0.069$). The results are qualitatively similar to our main results suggesting that natural disaster intensity has a positive effect on SE, and this effect is larger for men relative to women, those with low human capital, and those who fear failure.

*4.2.3 Alternative measures of our focal variable*

In our main analysis, we used the total number of affected people as the indicator of natural disaster intensity. This measure is the aggregation of the number of people that were affected, injured, homeless, and killed. As robustness checks, we use two alternative measures of natural disaster intensity—the number of affected people and the number of injured people (see the supplemental appendix). Table A2 reports the results using the number of people injured, and Table A3 reports the results using the number of people affected. Overall, the results are similar to our main findings. We find that natural disaster intensity has a positive effect on SE, and this effect is larger for men relative to women, those with low human capital, and those who are afraid of failure. These findings provide additional supports for our hypotheses.

*4.2.4 Additional controls for opportunity entrepreneurship rate and foreign aid*

To account for opportunities that emerged in the aftermath of natural disasters and the increased resources brought by international aid, we include opportunity entrepreneurship rate and foreign aid as additional control variables. Opportunity entrepreneurship rate measures the percentage of those involved in entrepreneurship that are opportunity motivated, we acquired this variable from GEM. Foreign aid is the total official development assistance received by recipient countries as a percentage of gross national investment (Boudreaux et al., 2022b). We gathered foreign aid data from WDI. We presented the results in the supplemental appendix Table A4, which are consistent with our main findings.

*4.2.5 Coarsened Exact Matching*

Although we attempted to control for potential issues of omitted variables and reverse causation using the instrumental variables approach, this method relies on the assumption that the instrument is uncorrelated with the residual and only influences the dependent variable through the endogenous regressor (Wooldridge, 2010). If this assumption fails, then the approach will lead to an inconsistent parameter estimate. Therefore, as an additional robustness check, we use the Coarsened Exact Matching (CEM) method to adjust our model for self-selection into SE. Unlike propensity score matching, CEM's advantage is that it does not estimate the probability of being treated (Nikolova et al., 2022). Rather, it coarsens the explanatory variables in strata and weights individuals based on their proximity to the treated group (Blackwell et al., 2009; Gustafsson et al., 2016; Iacus et al., 2012; King and Nielsen, 2016). We match SE based on several characteristics including age, entrepreneurial experience, gender, human capital, and fear of failure. For a match to be successful, we should observe a smaller $L1$ distance post-treatment (Blackwell et al., 2009; Iacus et al., 2012).

The $L1$ statistic for each variable as well as the multivariate $L1$ statistic is smaller post-treatment. This can be observed by comparing the supplemental appendix Tables A5 and A6. As such, we use the CEM weights in our analysis to adjust for the self-selection concern. We report the results of two estimators using CEM weights—a multi-level approach (random effects) and the country fixed effects approach. The supplemental appendix reports these results in Tables A7 and A8, respectively. Overall, the results are qualitatively similar to our main findings. We find that natural disaster intensity has a positive effect on SE, and this effect is larger for men relative to women, those with low human capital, and those who are afraid of failure. The parameter estimates are statistically significant for all direct effects of SE and moderators in Table A8 and statistically significant for all direct effects of SE and the fear of failure moderator in Table A7. In contrast, the results are similar in magnitude but statistically insignificant for the gender and human capital moderators in Table A7. Thus, the CEM results are consistent with our main findings, albeit weaker for some moderators than others. Nevertheless, the parameter estimates are qualitatively similar.

## 5 Discussion

Our objective was to provide theoretical and empirical insights into whether and how natural disasters affect SE. With respect to whether natural disasters affect SE, our analysis of 107,386 observations across 30 countries revealed that natural disaster intensity has a positive effect on individuals' likelihood to enter SE. We used ABV to understand how natural disasters affect SE. ABV highlights external effects on individuals' attention allocation rely on personal attributes (Ocasio et al., 2018). Our analysis revealed the positive effect of natural disaster intensity on SE is stronger for men, those with low human capital, and those who fear failure.

*5.1 Contributions to and implications for the SE literature*

Our results extend the SE literature. Drawing on ABV, this study investigates the joint effects of a novel contextual factor (i.e., natural disaster intensity) and personal attributes (i.e., gender, human capital, and fear of failure) on individuals' likelihood to enter SE, which develops a cross-level research model to draw a more comprehensive picture of the antecedents of SE. Since SE plays a pivotal role in addressing various social problems (Alvord et al., 2004; Bruton et al., 2021; Tobias et al., 2013), it is imperative to identify the factors driving individuals to engage in SE. The literature reveals both contextual and individual factors influence individuals' likelihood to engage in SE (Hockerts, 2017; Hota, 2021). Yet, despite the importance of these studies, they have progressed in isolation and fail to depict how these two levels of factors interactively affect the likelihood to pursue SE (Hechavarría and Brieger, 2022; Hockerts, 2017; Hoogendoorn, 2016; Miller et al., 2012; Pathak and Muralidharan, 2018; Stephan et al., 2015). Building on this literature, we adopt ABV, which suggests individuals' behaviors are jointly determined by the context they are embedded in and their personal attributes, to construct a cross-level model. We find natural disasters may stimulate individuals to allocate attention to social problems and then has a positive effect on the likelihood to pursue SE, and this positive effect is stronger for men, individuals lacking human capital, and those who fear failure. In this regard, this study not only indicates context and personal attributes jointly determine individuals' pursuit of SE but also lays a threshold over which to utilize ABV to explore the role of attention in SE. Hence, our study complements and extends the SE literature.

Future work might consider alternative boundary effects besides gender, human capital, and fear of failure that might direct individuals' attention toward SE in the aftermath of natural disasters or other catastrophes. One extension, for instance, is to consider the role of founder experience. Do founders with business experience react more quickly and efficiently to SE in the event of natural disasters? On the one hand, more experienced founders can leverage their industry specific knowledge and social capital to better navigate the uncertainty following natural disasters. On the other hand, more experienced founders might be more entrenched in their ways, resistant to change, and slower to adapt in the crisis. In addition, individuals may not respond to crisis alone. Instead, they can build teams and harness collective intelligence to decide their attention location. Hence, the characteristics of founding teams, such as team heterogeneity, are also worth further investigation, because these attributes affect a founding team's attention scope. In turn, this influences their response to natural disasters.

*5.2 Contributions to and implications for the natural disasters literature*

Our study also extends and has important implications for the natural disasters literature. By linking natural disasters to individuals' pursuit of SE and probing the boundaries of this linkage, we extend the consequences of natural disasters. While scholars have explored the effect of natural disasters on socioeconomic activities at country and regional levels (Boudreaux et al., 2022b, 2022c, 2019a; Dell et al., 2014), we still know little about the consequences of natural disasters at the individual level (Huang et al., 2018). Because natural disasters exert continuous threats to social members and serve as an inevitable determinant of individual behaviors, scholars advocate exploring the effects of natural disasters at the individual level (Huang et al., 2018). This study responds to this call by examining how natural disaster intensity influences SE pursuit. Importantly, our analysis reveals a key role of personal attributes—the positive relationship between natural disaster intensity and SE is stronger for men, those with low human capital, and those who fear failure. Insights like this cannot be examined by studies examining the effects of natural disasters on entrepreneurship at the country or regional-level.

Our study also extends the literature exploring the effect of natural disasters on entrepreneurship. Prior findings have been inconclusive. For example, while Boudreaux et al. (2022b, 2022c, 2019a) report that natural disasters increase uncertainty and then inhibit entrepreneurship, several case studies concluded natural disasters create market inefficiencies and new opportunities for entrepreneurs to exploit (Muñoz et al., 2019; Williams and Shepherd, 2016a, 2016b). A plausible explanation for this ambiguity is that these studies do not distinguish the effects of natural disasters on different types of entrepreneurship. For example, Boudreaux et al. (2022b, 2022c, 2019a) measured new venture creation by the number of newly registered firms, which is mainly based on general entrepreneurship. In contrast, the positive findings of Williams and Shepherd (2016a, 2016b) are more similar to SE, because the victims' entrepreneurial activities after disasters are commonly aiming at alleviating the social problems emerging from natural disasters. In this study, we focus on the effect of natural disasters on a specific type of entrepreneurship—SE, and we find that natural disaster intensity increases individuals' likelihood to engage in SE. This result provides additional empirical support for the findings of Williams and Shepherd (2016a, 2016b). As such, our study extends the consequences of natural disasters to individual social entrepreneurial behaviors.

We invite future research to consider distinguishing the role of natural disasters in distinct types of entrepreneurship. In doing so, future research can help unveil both the bright side (i.e., encouragement of SE) and dark side (i.e., inhibition of commercial entrepreneurship) of natural disasters, depending on the type of entrepreneurship. Studies adopting this nuanced perspective will help provide a more refined picture of how entrepreneurs respond to natural disasters, which will help facilitate managerial and policy discussions.

Lastly, our study contributes to the literature on entrepreneurs' responses to crises (Bullough et al., 2014; Bullough and Renko, 2017; Davidsson and Gordon, 2016; Doern, 2016; Doern et al., 2019). Although studies document negative effects of crises on entrepreneurship including failure, contraction, and resource destruction (Doern, 2016), the literature also identifies moderating factors with the potential to attenuate this negative impact like resilience (Bullough et al., 2014), social capital (Martinelli et al., 2018), prior experience (Doern, 2016), and commitment and ambition (Davidsson and Gordon, 2016). Our study complements this literature by documenting that personal attributes such as gender, human capital, and fear of failure can help spur individuals to enter SE.

*5.3 Managerial and policy implications*

Our study has important managerial implications for entrepreneurs. Our results reveal natural disaster intensity has a positive effect on individuals' pursuit of SE. Accordingly, instead of simply viewing natural disasters as sources of uncertainty and risk, individuals can take advantage of opportunities created by natural disasters to engage in SE and then address the challenges and adversities. In addition, our study finds that the positive effect of natural disaster intensity on individuals' likelihood to pursue SE is greater for men, individuals with low human capital, and with high fear of failure. Since these groups of persons pay less attention to SE-related issues and cannot rely on themselves to engage in SE, they may achieve their goals, such as creating social values, by using the opportunities provided by natural disasters.

Our study also offers suggestions for policymakers. Government and international aid organizations, though often well-intended, cannot fully address the adversities and social issues caused by natural disasters (Asongu, 2015; Asongu and Nwachukwu, 2017; Boudreaux et al., 2022b). Given that we find natural disasters' attention direction effects are more noticeable for men, individuals lacking human capital, and those who fear failure, policymakers might consider increasing media coverage of natural disasters to attract such persons' attention and then promote them to take actions, such as engaging in SE. Encouraging and supporting SE in the wake of natural disaster events will help to solve these social problems.

*5.4 Limitations and future directions*

Like any study, ours has several limitations that provide avenues for future studies. One limitation is that, although data from GEM have been widely used in previous studies, these data suffer limitations. For example, GEM measures SE by only adopting a single item. Hence, future studies can leverage other sources of data to further probe the influence of natural disaster intensity on SE. In addition, although our study examines the linkage between natural disaster intensity and individuals' pursuit of SE, our measure of SE might primarily reflect the quantity of SE and offer little insight on the quality of SE. Future studies, thus, might explore the influence of natural disaster intensity on the quality of SE. Third, while ABV has been an important theoretical lens to help explain organizational decision-making and actions, its application to SE remains in infancy. Thus, future studies can build on our study to further explore how other attention-related factors affect individuals' attention to social problems and their pursuit of SE. Fourth, this study adopts the number of affected people to capture natural disaster intensity and treats it as homogeneous for a country. Natural disaster intensity varies strongly among regions (i.e., coastal areas vs. non-coastal areas and cities in seismic zones versus cities in non-seismic zones). Hence, future studies can use the regional-level data to permit a better examination of natural disasters' effects. Lastly, this study examines the effect of natural disasters on SE by utilizing cross-section data. Future studies can adopt panel data and quasi-experimental methods such as synthetic control to further validate the positive effect of natural disasters on SE.

In addition, since ABV stresses the joint effect of contextual and individual attributes in attention allocation, and subsequent decision and behavior, the application of ABV in SE validates the explanatory power of ABV as a theory. Furthermore, the application of ABV as an emerging theoretical lens in other contexts (e.g., other type of entrepreneurship) may contribute to ABV itself since it may or may not explain other type of entrepreneurship. In so doing, we can test, apply, or even modify ABV in the domain of entrepreneurship.

## 6 Conclusion

This study examines how natural disasters influence SE. Drawing on data from 107,386 observations across 30 countries in 2015, we find that natural disaster intensity has a positive effect on individuals' likelihood to engage in SE. Moreover, this effect is greater for men, individuals with a low level of human capital, and individuals with high fear of failure. The results suggest that both contextual and individual factors are key determinants of SE, which contributes to drawing a more comprehensive picture of the antecedents of SE. Moreover, it also enriches our knowledge of the implications of natural disasters for individual behaviors.

Loayza, N.V., Olaberría, E., Rigolini, J., Christiaensen, L., 2012. Natural Disasters and Growth: Going Beyond the Averages. World Development 40, 1317–1336. https://doi.org/10.1016/j.worlddev.2012.03.002

Maki, A., Dwyer, P.C., Blazek, S., Snyder, M., González, R., Lay, S., 2019. Responding to natural disasters: Examining identity and prosociality in the context of a major earthquake. British Journal of Social Psychology 58, 66–87. https://doi.org/10.1111/bjso.12281

Marino, L.D., Lohrke, F.T., Hill, J.S., Weaver, K.M., Tambunan, T., 2008. Environmental Shocks and SME Alliance Formation Intentions in an Emerging Economy: Evidence from the Asian Financial Crisis in Indonesia. Entrepreneurship Theory and Practice 32, 157–183. https://doi.org/10.1111/j.1540-6520.2007.00220.x

Martinelli, E., Tagliazucchi, G., Marchi, G., 2018. The resilient retail entrepreneur: dynamic capabilities for facing natural disasters. International Journal of Entrepreneurial Behavior & Research 24, 1222–1243. https://doi.org/10.1108/IJEBR-11-2016-0386

Mason, C., Brown, R., 2013. Creating good public policy to support high-growth firms. Small Business Economics 40, 211–225. https://doi.org/10.1007/s11187-011-9369-9

Mazzucato, M., 2018. Mission-oriented innovation policies: challenges and opportunities. Industrial and Corporate Change 27, 803–815. https://doi.org/10.1093/icc/dty034

Miller, T.L., Grimes, M.G., McMullen, J.S., Vogus, T.J., 2012. Venturing for Others with Heart and Head: How Compassion Encourages Social Entrepreneurship. Academy of Management Review 37, 616–640. https://doi.org/10.5465/amr.2010.0456

Monllor, J., Murphy, P.J., 2017. Natural disasters, entrepreneurship, and creation after destruction: A conceptual approach. International Journal of Entrepreneurial Behavior & Research 23, 618–637. https://doi.org/10.1108/IJEBR-02-2016-0050

Morgan, J., Sisak, D., 2016. Aspiring to succeed: A model of entrepreneurship and fear of failure. Journal of Business Venturing 31, 1–21. https://doi.org/10.1016/j.jbusvent.2015.09.002

Muñoz, P., Kimmitt, J., Kibler, E., Farny, S., 2019. Living on the slopes: entrepreneurial preparedness in a context under continuous threat. Entrepreneurship & Regional Development 31, 413–434. https://doi.org/10.1080/08985626.2018.1541591

Nadkarni, S., Barr, P.S., 2008. Environmental context, managerial cognition, and strategic action: an integrated view. Strategic Management Journal 29, 1395–1427. https://doi.org/10.1002/smj.717

Nicolás, C., Rubio, A., Fernández-Laviada, A., 2018. Cognitive Determinants of Social Entrepreneurship: Variations According to the Degree of Economic Development. Journal of Social Entrepreneurship 9, 154–168. https://doi.org/10.1080/19420676.2018.1452280

Nikolova, M., Nikolaev, B., Boudreaux, C., 2022. Being your own boss and bossing others: the moderating effect of managing others on work meaning and autonomy for the self-employed and employees. Small Business Economics. https://doi.org/10.1007/s11187-021-00597-z

Ocasio, W., 1997. Towards an Attention-Based View of the Firm. Strategic Management Journal 18, 187–206. https://doi.org/10.1002/(SICI)1097-0266(199707)18:1+<187::AID-SMJ936>3.0.CO;2-K

Ocasio, W., Laamanen, T., Vaara, E., 2018. Communication and attention dynamics: An attention-based view of strategic change. Strategic Management Journal 39, 155–167. https://doi.org/10.1002/smj.2702

Parker, S.C., 2018. The economics of entrepreneurship. Cambridge University Press.

Pathak, S., Muralidharan, E., 2018. Economic Inequality and Social Entrepreneurship. Business & Society 57,
35

1150–1190. https://doi.org/10.1177/0007650317696069

Pinkse, J., Gasbarro, F., 2019. Managing Physical Impacts of Climate Change: An Attentional Perspective on Corporate Adaptation. Business & Society 58, 333–368. https://doi.org/10.1177/0007650316648688

Rieger, V., Gründler, A., Winkler, H.-J., Tschauner, B., Engelen, A., 2021. A cross-national perspective of compassion's role in driving social entrepreneurial intentions. Journal of International Management 27, 100824. https://doi.org/10.1016/j.intman.2021.100824

Robinson, W., 1950. Ecological Correlations and the Behavior of Individuals. American Sociological Review 15, 351–357.

Ruskin, J., Seymour, R.G, Webster, C.M., 2016. Why Create Value for Others? An Exploration of Social Entrepreneurial Motives. Journal of Small Business Management 54, 1015–1037. https://doi.org/10.1111/jsbm.12229

Saebi, T., Foss, N.J., Linder, S., 2019. Social Entrepreneurship Research: Past Achievements and Future Promises. Journal of Management 45, 70–95. https://doi.org/10.1177/0149206318793196

Sahasranamam, S., Nandakumar, M.K., 2020. Individual capital and social entrepreneurship: Role of formal institutions. Journal of Business Research 107, 104–117. https://doi.org/10.1016/j.jbusres.2018.09.005

Salvato, C., Sargiacomo, M., Amore, M.D., Minichilli, A., 2020. Natural disasters as a source of entrepreneurial opportunity: Family business resilience after an earthquake. Strategic Entrepreneurship Journal 14, 594–615. https://doi.org/10.1002/sej.1368

Santos, F.M., 2012. A Positive Theory of Social Entrepreneurship. Journal of Business Ethics 111, 335–351. https://doi.org/10.1007/s10551-012-1413-4

Schofer, E., Fourcade-Gourinchas, M., 2001. The Structural Contexts of Civic Engagement: Voluntary Association Membership in Comparative Perspective. American Sociological Review 66, 806–828. https://doi.org/10.2307/3088874

Seligson, M.A., 2002. The Renaissance of Political Culture or the Renaissance of the Ecological Fallacy? Comparative Politics 34, 273–292. https://doi.org/10.2307/4146954

Shabnam, N., 2014. Natural Disasters and Economic Growth: A Review. International Journal of Disaster Risk Science 5, 157–163. https://doi.org/10.1007/s13753-014-0022-5

Shane, S., 2009. Why encouraging more people to become entrepreneurs is bad public policy. Small Business Economics 33, 141–149. https://doi.org/10.1007/s11187-009-9215-5

Shepherd, D.A., Williams, T.A., 2014. Local Venturing as Compassion Organizing in the Aftermath of a Natural Disaster: The Role of Localness and Community in Reducing Suffering. Journal of Management Studies 51, 952–994. https://doi.org/10.1111/joms.12084

Skidmore, M., Toya, H., 2002. Do Natural Disasters Promote Long-Run Growth? Economic Inquiry 40, 664–687. https://doi.org/10.1093/ei/40.4.664

Stephan, U., Uhlaner, L.M., Stride, C., 2015. Institutions and social entrepreneurship: The role of institutional voids, institutional support, and institutional configurations. Journal of International Business Studies 46, 308–331. https://doi.org/10.1057/jibs.2014.38

Stevens, R., Moray, N., Bruneel, J., Clarysse, B., 2015. Attention allocation to multiple goals: The case of for-profit social enterprises. Strategic Management Journal 36, 1006–1016. https://doi.org/10.1002/smj.2265

Su, Z., Yang, J., Wang, Q., 2022. The Effects of Top Management Team Heterogeneity and Shared Vision on
36

**Table 1** Distribution of observations among countries (N=107,386)

| Country | N | Percentage | Country | N | Percentage |
| --- | --- | --- | --- | --- | --- |
| Argentina | 2,951 | 2.75% | Mexico | 4,113 | 3.83% |
| Australia | 1,894 | 1.76% | Morocco | 1,953 | 1.82% |
| Brazil | 1,981 | 1.84% | Peru | 1,951 | 1.82% |
| Chile | 6,132 | 5.71% | Philippines | 1,949 | 1.81% |
| China | 3,545 | 3.30% | Poland | 1,735 | 1.62% |
| Colombia | 3,652 | 3.40% | Romania | 1,901 | 1.77% |
| Ecuador | 2,099 | 1.95% | Slovenia | 1,911 | 1.78% |
| Germany | 3,742 | 3.48% | South Africa | 3,016 | 2.81% |
| Guatemala | 2,156 | 2.01% | South Korea | 1,891 | 1.76% |
| Hungary | 1,901 | 1.77% | Spain | 23,664 | 22.04% |
| India | 3,149 | 2.93% | Switzerland | 2,270 | 2.11% |
| Indonesia | 5,251 | 4.89% | Thailand | 2,982 | 2.78% |
| Iran | 2,988 | 2.78% | United Kingdom | 9,002 | 8.38% |
| Italy | 1,933 | 1.80% | United States | 2,333 | 2.17% |
| Kazakhstan | 1,424 | 1.33% | Vietnam | 1917 | 1.79% |

**Table 2** Measures of variables and their sources

| Variables | Measures | Sources |
|---|---|---|
| Social entrepreneurship (SE) | 1 = respondent involved in a social startup, 0 otherwise. | GEM |
| Gender | 1 = male, 0 = female | GEM |
| Human capital | 1 = respondent has a college education or above, 0 otherwise | GEM |
| Fear of failure | 1 = respondent would not start a business out of fear of failure, 0 otherwise. | GEM |
| Age | Age in years (linear and squared). | GEM |
| Entrepreneurial experience | Respondent sold, shut down, discontinued, or quit a business in the past 12 months that she/he owned and managed, and this business continued its activities after the entrepreneur disengaged. | GEM |
| Unemployment rate | The percentage of unemployed population in the economically active population. | WDI |
| GDP per capita | Gross Domestic Product (GDP) per capita (in *ln* form). | WDI |
| Human inequality | Coefficient of human inequality in human development report (calculated as the mean of the values in inequality in life expectancy, education, and income). | UN |
| Postmaterialism | The percentage of individuals in each country's sample that were scored as postmaterialists. | WVS |
| Governance quality | The average of six indicators (voice and accountability, political stability, government effectiveness, regulatory quality, rule of law, and control of corruption). | WGI |
| Government spending | The consumption and all transfer payments by the state. | EFI |
| Fiscal health | Measured by two sub-factors: average deficits as a percentage of GDP for the most recent three years and debt as a percentage of GDP. | EFI |
| Disaster intensity(*ln*) | The natural logarithm of the total affected people in natural disasters. | CRED |

**Table 3** Descriptive statistics and correlation matrix

| Variables | 1 | 2 | 3 | 4 | 5 | 6 | 7 | 8 | 9 | 10 | 11 | 12 | 13 | 14 |
|---|---|---|---|---|---|---|---|---|---|---|---|---|---|---|
| 1. SE | 1 | | | | | | | | | | | | | |
| 2. Gender | 0.027 | 1 | | | | | | | | | | | | |
| 3. Human capital | 0.088 | 0.014 | 1 | | | | | | | | | | | |
| 4. Fear of fail | -0.028 | -0.069 | -0.004 | 1 | | | | | | | | | | |
| 5. Age | -0.001 | -0.020 | -0.017 | -0.016 | 1 | | | | | | | | | |
| 6. Entrepreneurial experience | 0.069 | 0.010 | 0.014 | -0.013 | 0.003 | 1 | | | | | | | | |
| 7. Unemployment rate | -0.085 | 0.011 | -0.089 | 0.025 | 0.064 | -0.054 | 1 | | | | | | | |
| 8. GDP per capita | -0.030 | 0.008 | 0.065 | 0.017 | 0.189 | -0.061 | 0.415 | 1 | | | | | | |
| 9. Human inequality | 0.024 | -0.001 | -0.083 | -0.054 | -0.164 | 0.069 | -0.208 | -0.750 | 1 | | | | | |
| 10. Postmaterialism | 0.021 | -0.009 | 0.067 | -0.042 | 0.097 | 0.009 | -0.197 | 0.499 | -0.306 | 1 | | | | |
| 11. Governance quality | 0.001 | 0.004 | 0.076 | 0.001 | 0.188 | -0.052 | 0.252 | 0.860 | -0.737 | 0.520 | 1 | | | |
| 12. Government spending | 0.055 | -0.008 | -0.046 | -0.033 | -0.143 | 0.077 | -0.499 | -0.720 | 0.702 | -0.322 | -0.607 | 1 | | |
| 13. Fiscal health | 0.068 | -0.009 | 0.041 | 0.009 | -0.016 | 0.045 | -0.455 | -0.071 | 0.028 | 0.118 | -0.005 | 0.481 | 1 | |
| 14. Disaster intensity (*ln*) | 0.095 | -0.010 | 0.011 | -0.050 | -0.106 | 0.070 | -0.699 | -0.624 | 0.586 | -0.102 | -0.494 | 0.651 | 0.223 | 1 |
| Mean | 0.059 | 0.494 | 0.221 | 0.418 | 41.032 | 0.037 | 10.415 | 9.530 | 16.039 | 11.547 | 0.432 | 58.575 | 68.706 | 8.231 |
| Standard deviation | 0.235 | 0.500 | 0.415 | 0.493 | 14.327 | 0.190 | 8.389 | 0.998 | 7.408 | 6.691 | 0.752 | 22.524 | 22.040 | 5.902 |

Note: Correlations > |0.005| are statistically significant at the 0.01 level.

**Table 4** The multilevel logistic regression results

| Variables | Model 1.1 | Model 1.2 | Model 1.3 | Model 1.4 | Model 1.5 | Model 1.6 |
|---|---|---|---|---|---|---|
| Age | 0.002 | 0.002 | 0.002 | 0.001 | 0.002 | 0.0001 |
| | (0.014) | (0.014) | (0.014) | (0.014) | (0.014) | (0.014) |
| Age square | -0.034** | -0.034** | -0.034** | -0.033** | -0.034** | -0.034** |
| | (0.012) | (0.012) | (0.012) | (0.012) | (0.012) | (0.012) |
| Entrepreneurial experience | 0.145*** | 0.145*** | 0.145*** | 0.145*** | 0.145*** | 0.145*** |
| | (0.009) | (0.009) | (0.009) | (0.009) | (0.009) | (0.009) |
| Gender | 0.115*** | 0.115*** | 0.103*** | 0.116*** | 0.115*** | 0.101*** |
| | (0.013) | (0.013) | (0.015) | (0.013) | (0.013) | (0.015) |
| Human capital | 0.278*** | 0.278*** | 0.278*** | 0.291*** | 0.278*** | 0.290*** |
| | (0.012) | (0.012) | (0.012) | (0.013) | (0.012) | (0.013) |
| Fear of failure | -0.072*** | -0.072*** | -0.072*** | -0.072*** | -0.090*** | -0.091*** |
| | (0.014) | (0.014) | (0.014) | (0.014) | (0.015) | (0.015) |
| Unemployment rate | -0.173 | 0.016 | 0.017 | 0.020 | 0.018 | 0.023 |
| | (0.232) | (0.215) | (0.215) | (0.215) | (0.216) | (0.216) |
| GDP per capita | -0.124 | -0.137 | -0.138 | -0.134 | -0.141 | -0.140 |
| | (0.243) | (0.214) | (0.214) | (0.214) | (0.215) | (0.215) |
| Human inequality | 0.106 | -0.021 | -0.022 | -0.020 | -0.023 | -0.023 |
| | (0.211) | (0.192) | (0.192) | (0.191) | (0.193) | (0.192) |
| Postmaterialism | -0.007 | 0.051 | 0.051 | 0.050 | 0.050 | 0.050 |
| | (0.148) | (0.132) | (0.132) | (0.132) | (0.133) | (0.133) |
| Governance quality | 0.189 | 0.194 | 0.194 | 0.193 | 0.196 | 0.195 |
| | (0.226) | (0.199) | (0.199) | (0.199) | (0.200) | (0.200) |
| Government spending | -0.117 | -0.219 | -0.219 | -0.220 | -0.220 | -0.220 |
| | (0.221) | (0.199) | (0.199) | (0.198) | (0.199) | (0.199) |
| Fiscal health | 0.169 | 0.259* | 0.259* | 0.259* | 0.260* | 0.261* |
| | (0.143) | (0.130) | (0.130) | (0.130) | (0.130) | (0.130) |
| Disaster intensity (DI) | | 0.477** | 0.474** | 0.492** | 0.480** | 0.491** |
| | | (0.163) | (0.163) | (0.163) | (0.164) | (0.164) |
| DI × Gender | | | 0.034* | | | 0.039* |
| | | | (0.016) | | | (0.016) |
| DI × Human capital | | | | -0.039** | | -0.040** |
| | | | | (0.014) | | (0.014) |
| DI × Fear of failure | | | | | 0.051** | 0.053** |
| | | | | | (0.016) | (0.016) |
| *Number of observations* | 107,386 | 107,386 | 107,386 | 107,386 | 107,386 | 107,386 |
| *Wald Chi-square* | 926.03*** | 935.98*** | 940.13*** | 944.24*** | 945.34*** | 958.87*** |
| *Likelihood ratio test* | 1800.88*** | 1365.15*** | 1365.25*** | 1359.84*** | 1373.29*** | 1368.51*** |
| *Log Likelihood* | -22005.54 | -22001.78 | -21999.48 | -21997.98 | -21996.84 | -21990.17 |

Note: Dependent variable is SE. Standard errors reported in parentheses.
$^+ p < 0.10$; $^* p < 0.05$; $^{**} p < 0.01$; $^{***} p < 0.001$ (Two-tailed tests).

**Table 5** Instrumental variable regression results

| Variables | First-stage<br>Disaster intensity | Second-stage: SE | | | | |
|---|---|---|---|---|---|---|
| | | Model 2.1 | Model 2.2 | Model 2.3 | Model 2.4 | Model 2.5 |
| Age | 0.008*** | 0.002 | -0.003 | -0.007 | -0.001 | -0.011+ |
| | (0.002) | (0.006) | (0.006) | (0.006) | (0.006) | (0.006) |
| Age square | 0.003* | -0.013* | -0.012* | -0.005 | -0.015** | -0.006 |
| | (0.001) | (0.005) | (0.005) | (0.005) | (0.005) | (0.005) |
| Entrepreneurial experience | 0.004** | 0.072*** | 0.069*** | 0.069*** | 0.067*** | 0.065*** |
| | (0.002) | (0.005) | (0.005) | (0.005) | (0.005) | (0.005) |
| Gender | -0.005** | 0.051*** | 0.045*** | 0.046*** | 0.046*** | 0.038*** |
| | (0.002) | (0.006) | (0.006) | (0.006) | (0.006) | (0.006) |
| Human capital | -0.007*** | 0.136*** | 0.130*** | 0.137*** | 0.127*** | 0.128*** |
| | (0.002) | (0.006) | (0.007) | (0.007) | (0.007) | (0.008) |
| Fear of failure | -0.009*** | -0.031*** | -0.033*** | -0.028*** | -0.025*** | -0.026*** |
| | (0.002) | (0.006) | (0.006) | (0.006) | (0.007) | (0.006) |
| Unemployment rate | -0.690*** | 0.505*** | 0.485*** | 0.391*** | 0.490*** | 0.380*** |
| | (0.003) | (0.041) | (0.041) | (0.047) | (0.042) | (0.045) |
| GDP per capita | -0.087*** | -0.089*** | -0.095*** | -0.045** | -0.110*** | -0.071*** |
| | (0.004) | (0.015) | (0.014) | (0.017) | (0.015) | (0.018) |
| Human inequality | 0.374*** | -0.299*** | -0.292*** | -0.212*** | -0.287*** | -0.206*** |
| | (0.003) | (0.027) | (0.027) | (0.032) | (0.028) | (0.030) |
| Postmaterialism | -0.091*** | 0.089*** | 0.090*** | 0.057*** | 0.089*** | 0.062*** |
| | (0.002) | (0.011) | (0.011) | (0.012) | (0.011) | (0.012) |
| Governance quality | 0.169*** | 0.023 | 0.024 | 0.032* | 0.046** | 0.051*** |
| | (0.004) | (0.016) | (0.015) | (0.015) | (0.016) | (0.015) |
| Government spending | 0.126*** | -0.204*** | -0.191*** | -0.149*** | -0.188*** | -0.132*** |
| | (0.004) | (0.016) | (0.017) | (0.020) | (0.018) | (0.019) |
| Fiscal health | -0.162*** | 0.224*** | 0.216*** | 0.164*** | 0.213*** | 0.156*** |
| | (0.002) | (0.012) | (0.012) | (0.018) | (0.013) | (0.018) |
| Population density | -0.112*** | | | | | |
| | (0.002) | | | | | |
| Disaster intensity (DI) | | 1.010*** | 0.956*** | 0.766*** | 0.953*** | 0.712*** |
| | | (0.056) | (0.061) | (0.079) | (0.064) | (0.076) |
| DI × Gender | | | 0.304** | | | 0.216* |
| | | | (0.098) | | | (0.098) |
| DI × Human capital | | | | -0.503*** | | -0.480*** |
| | | | | (0.084) | | (0.083) |
| DI × Fear of failure | | | | | 0.329*** | 0.273** |
| | | | | | (0.095) | (0.094) |
| *Number of observations* | 107,386 | 107,386 | 107,386 | 107,386 | 107,386 | 107,386 |
| *Wald Chi-square* | 3602.39*** | 3797.80*** | 4393.62*** | 4342.31*** | 4409.11*** | 5142.02*** |
| *First stage F-statistic* | 19833.02*** | - | - | - | - | - |
| *Pseudo likelihood* | - | -106403.79 | -258526.65 | -253012.43 | -258662.44 | -557069.78 |

Note: Dependent variable is SE. Standard errors are reported in parentheses.
+ $p < 0.10$; * $p < 0.05$; ** $p < 0.01$; *** $p < 0.001$ (Two-tailed tests).

**Figure 1.** The conceptual model

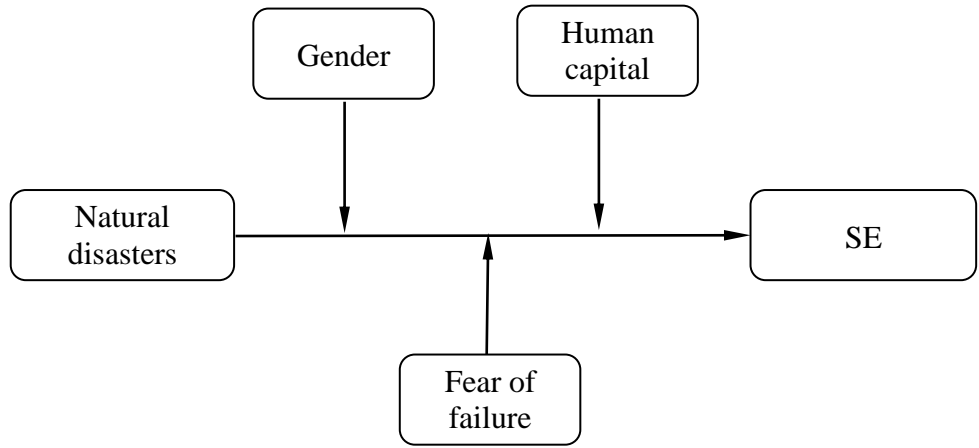

**Figure 2.** The interactive effect of natural disaster intensity and gender on SE

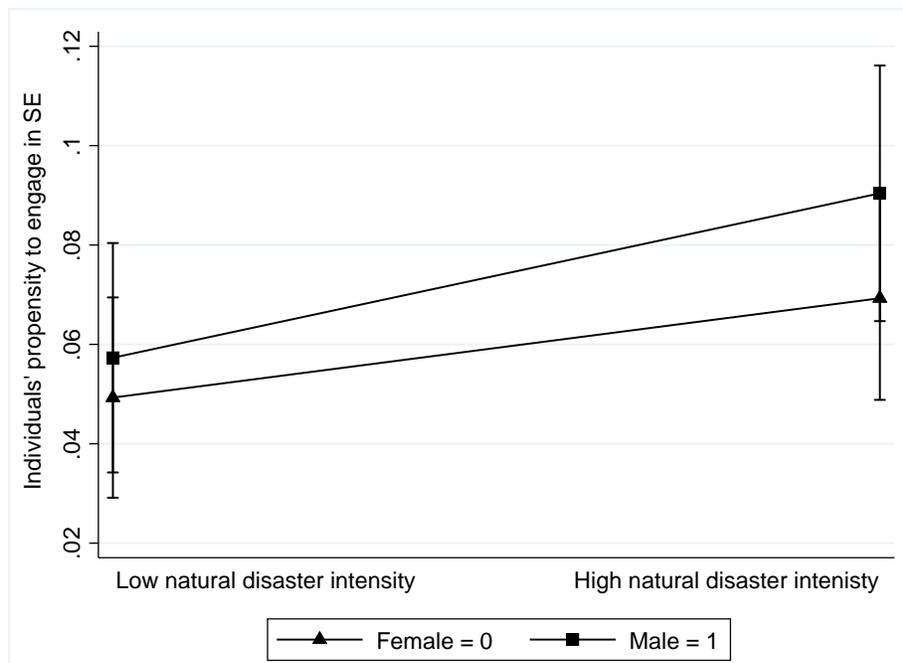

**Figure 3.** The interactive effect natural disaster intensity and human capital on SE

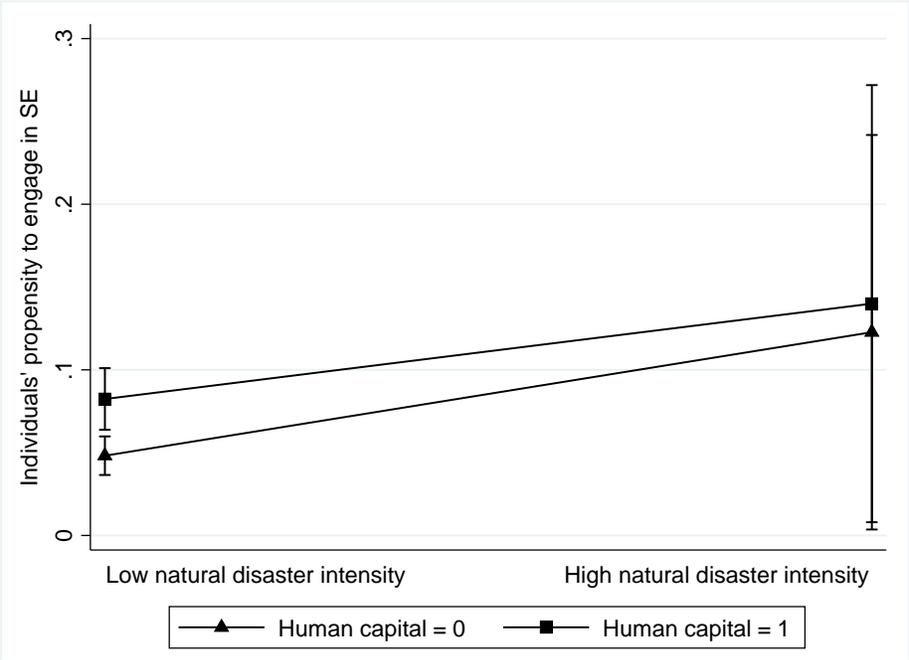

**Figure 4.** The interactive effect of natural disaster intensity and fear of failure on SE

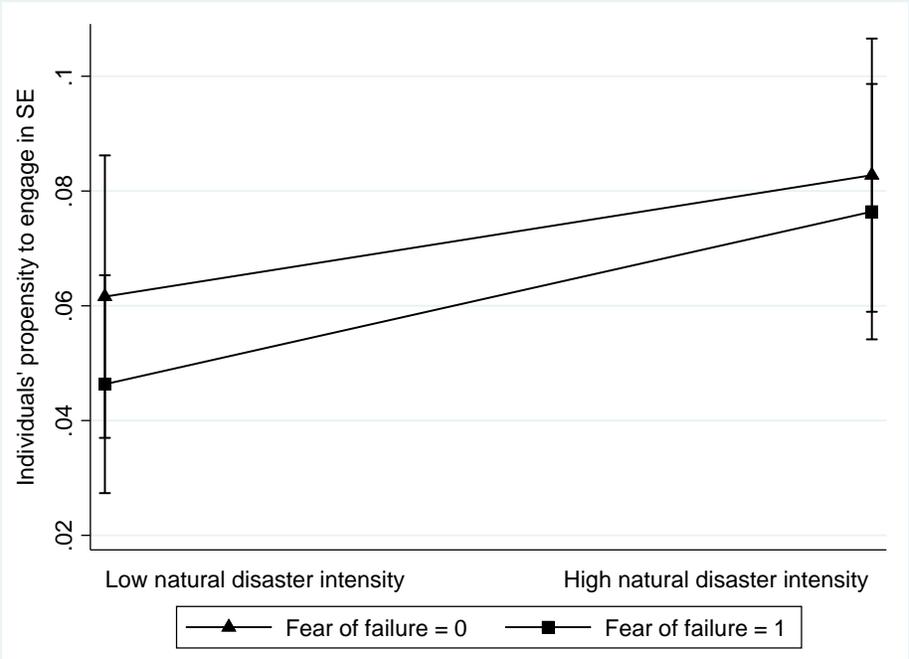